\documentclass[aps,prl,twocolumn,superscriptaddress]{revtex4}
\usepackage{graphicx}
\usepackage{subfigure}
\usepackage[usenames]{color}
\begin{document}
\bibliographystyle{unsrt}
\author{Michael Foss-Feig}
\affiliation{
Department of Physics, University of Colorado, Boulder, Colorado 80309, USA}
\affiliation{ 
JILA, Boulder, Colorado 80309, USA}
\author{Michael Hermele}
\affiliation{
Department of Physics, University of Colorado, Boulder, Colorado 80309, USA}
\author{Ana Maria Rey}
\affiliation{
Department of Physics, University of Colorado, Boulder, Colorado 80309, USA}
\affiliation{ 
JILA, Boulder, Colorado 80309, USA}
\affiliation{NIST, Boulder, Colorado 80309, USA}
\title{Probing the Kondo Lattice Model with Alkaline Earth Atoms}
\date{\today}
\begin{abstract}
We study transport properties of alkaline-earth atoms governed by the
Kondo Lattice Hamiltonian plus a harmonic confining potential, and suggest simple dynamical probes of
several different regimes of the phase diagram that can be
implemented with current experimental techniques.  In particular, we
show how Kondo physics at strong coupling, low density,
and in the heavy fermion phase is manifest in the dipole oscillations
of the conduction band upon displacement of the trap center.
\end{abstract}
\maketitle
\setlength{\parskip}{0pt}
\setlength{\abovecaptionskip}{0pt}
\setlength{\belowcaptionskip}{0pt}

To date, most cold atom simulations of condensed matter systems have
focused on the single-band Bose and Fermi Hubbard models, both because they are relatively simple to simulate and because they are believed to capture a great deal of important physics.  However, there are many real materials
in which the relevance of both internal (spin) and external (orbital) electronic degrees of
freedom preclude description by the single-band Hubbard model.  Recently it has
been shown that fermionic alkaline-earth atoms (AEAs) have unique properties
that allow for simulation of Hamiltonians with both spin and orbital
degrees of freedom \cite{Gorshkov:2009p4747}.  Here we discuss avenues thereby opened into optical lattice simulations of the Kondo Lattice Model (KLM).  As is generally the case, the necessity to perform cold atom simulation in a trap complicates the analogy with the translationally invariant KLM. 
 However, in this paper we
emphasize how a trap can help reveal hallmarks of the KLM, including heavy fermion mass enhancement
(through dynamics induced by trap displacement),
and the Kondo insulator gap (through formation
of a density plateau).  The proposed experimental probing methods (center of mass oscillations \cite{Fertig:2005p4167,Strohmaier2007,McKay2008} and shell structure density profiles \cite{Campbell:2006p5171,Folling:2006p5135}) have  been demonstrated to be successful diagnostic tools  in alkali atoms and therefore we expect that our analysis  will have direct applicability in near future experiments done with alkaline-earth atoms.

In its standard form, the KLM consists of a band of conduction electrons
interacting via a contact Heisenberg exchange with a lattice of immobile spins.  We focus on the case of antiferromagnetic (AF) exchange, relevant to so-called heavy fermion materials,  known for radically enhanced quasiparticle masses \cite{Doniach:1977}.  Simulation of the KLM Hamiltonian with AEAs was described in \cite{Gorshkov:2009p4747}, and here we only summarize the basic idea.

The $^1S_0$ ($g$) and $^3P_0$ ($e$) clock states of an AEA can be
trapped independently in two different  optical
lattice potentials  whose periodicities could be engineered to be the same \cite{Daley:2008p5398}.  We can therefore consider  a Mott insulator of $e$ atoms (immobile spins) trapped in a deep optical lattice and  coexisting with mobile $g$ atoms (conduction electrons) trapped in a shallow lattice with the same periodicity.  At low temperatures the interactions are determined by  4 s-wave scattering lengths $a_{ee}$, $a_{gg}$ and $a_{eg}^{\pm}$ for the states $\left|ee\right>$, $\left|gg\right>$ and $\frac{1}{\sqrt{2}}\left(\left|eg\right>\pm\left|ge\right>\right)$ respectively.  We choose the $e$ atoms to be localized because they would otherwise suffer lossy collisions.  Collisions between $g$ and $e$ atoms on the other hand are expected to be mostly elastic \cite{Gorshkov:2009p4747}.   The independence of the scattering lengths on the nuclear spin state guarantees that there will not be spin changing collisions, and so we are justified to consider an ensemble with only two nuclear spin states $\sigma=\pm$ (the electron spin in the KLM).
\begin{figure}[t]
\centering
\subfiguretopcaptrue
\subfigure[][]{
\label{PD1D}
\includegraphics[width=3.57 cm]{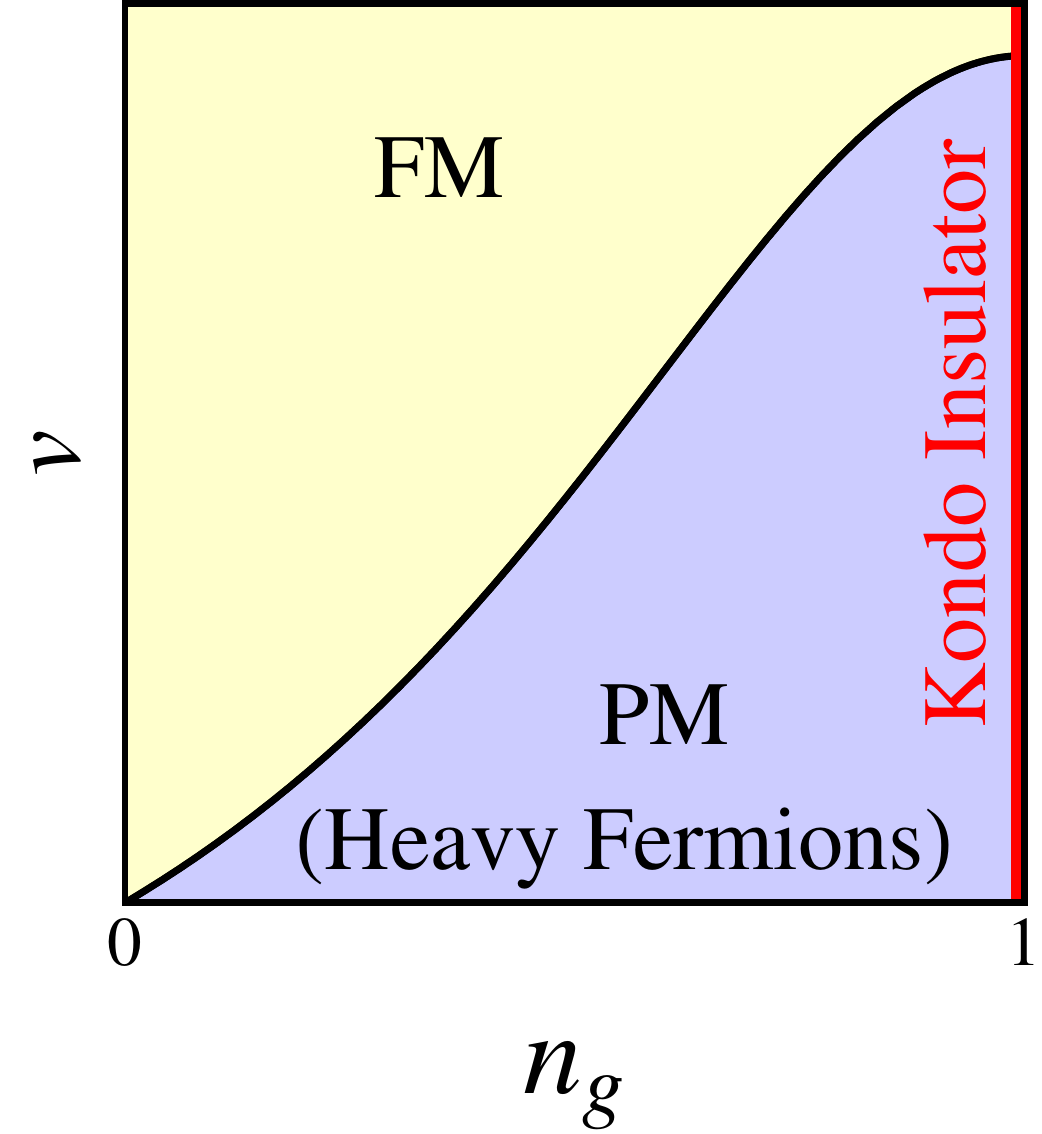}
}
\subfigure[][]{
\label{latticegraphics}
\includegraphics[width=4.54 cm]{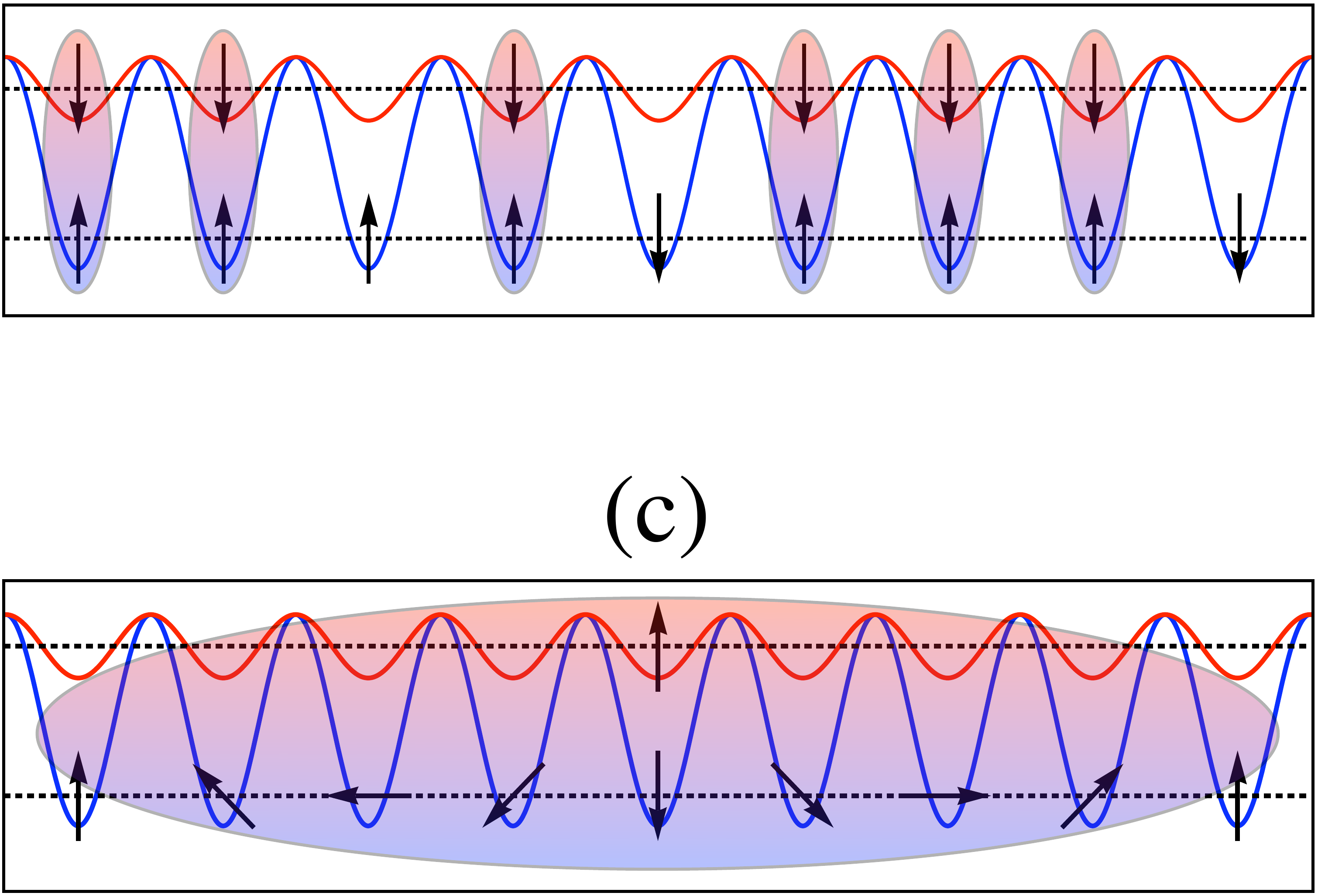}
\vspace{-1cm}}
\caption{(Color Online) \subref{PD1D} Schematic zero temperature phase diagram for the 1-D KLM \cite{Tsunetsugu:1997p597,GulX000E1Csi:2004p148}.  FM is a ferromagnetic phase, and PM is a paramagnetic phase closely related to heavy fermions in higher dimensions.  Schematic of the ground state at strong coupling (b) and for one $g$ atom (c).}
\label{phasediagrams}
\end{figure}
\begin{figure*}[!t]
\centering
\subfiguretopcaptrue

\subfigure[][]{
\hspace{-.1in}
\label{phyb}
\includegraphics[height=3.5 cm]{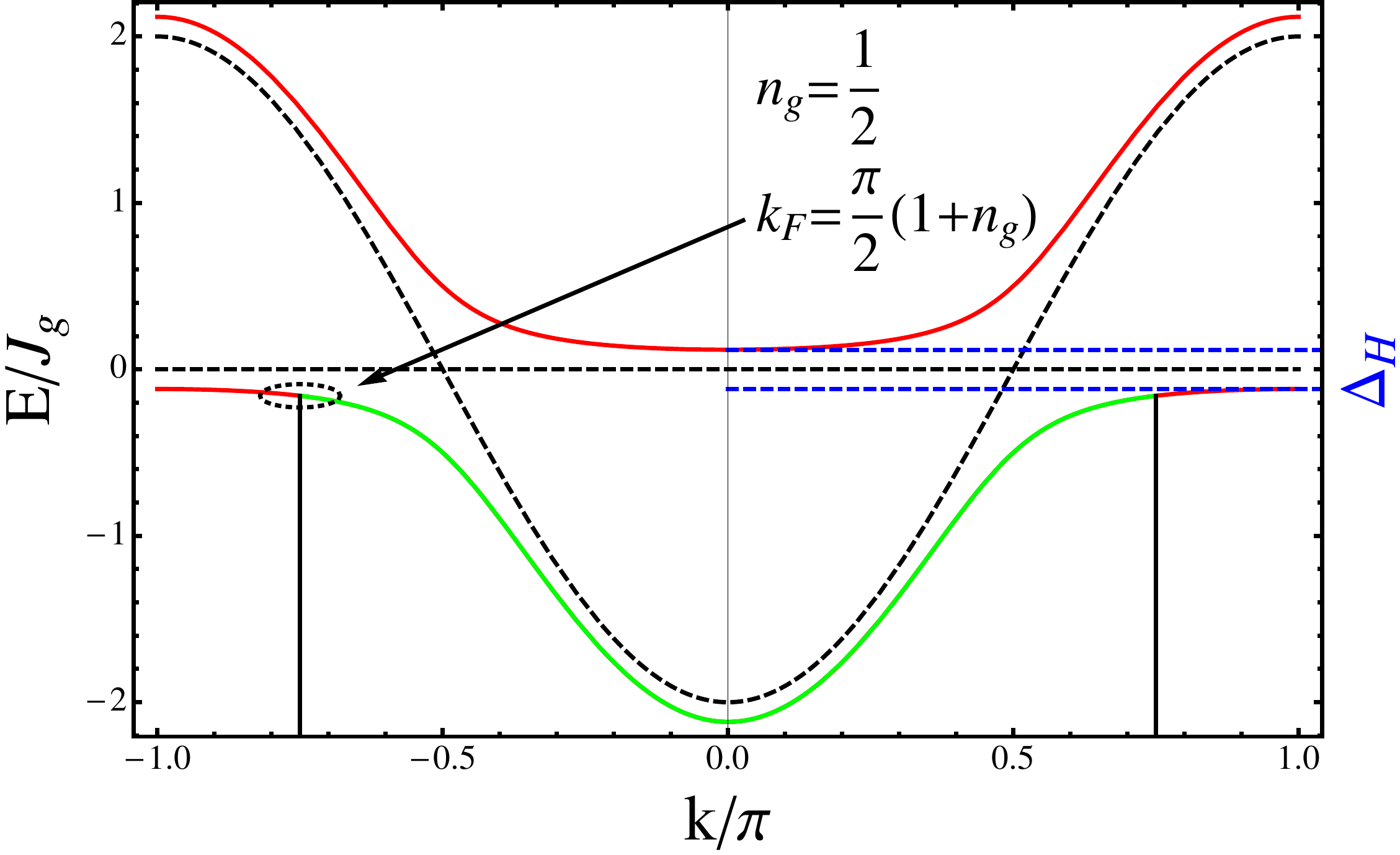}
}
\subfigure[][]{
\label{hybplateau}
\includegraphics[height=3.5 cm]{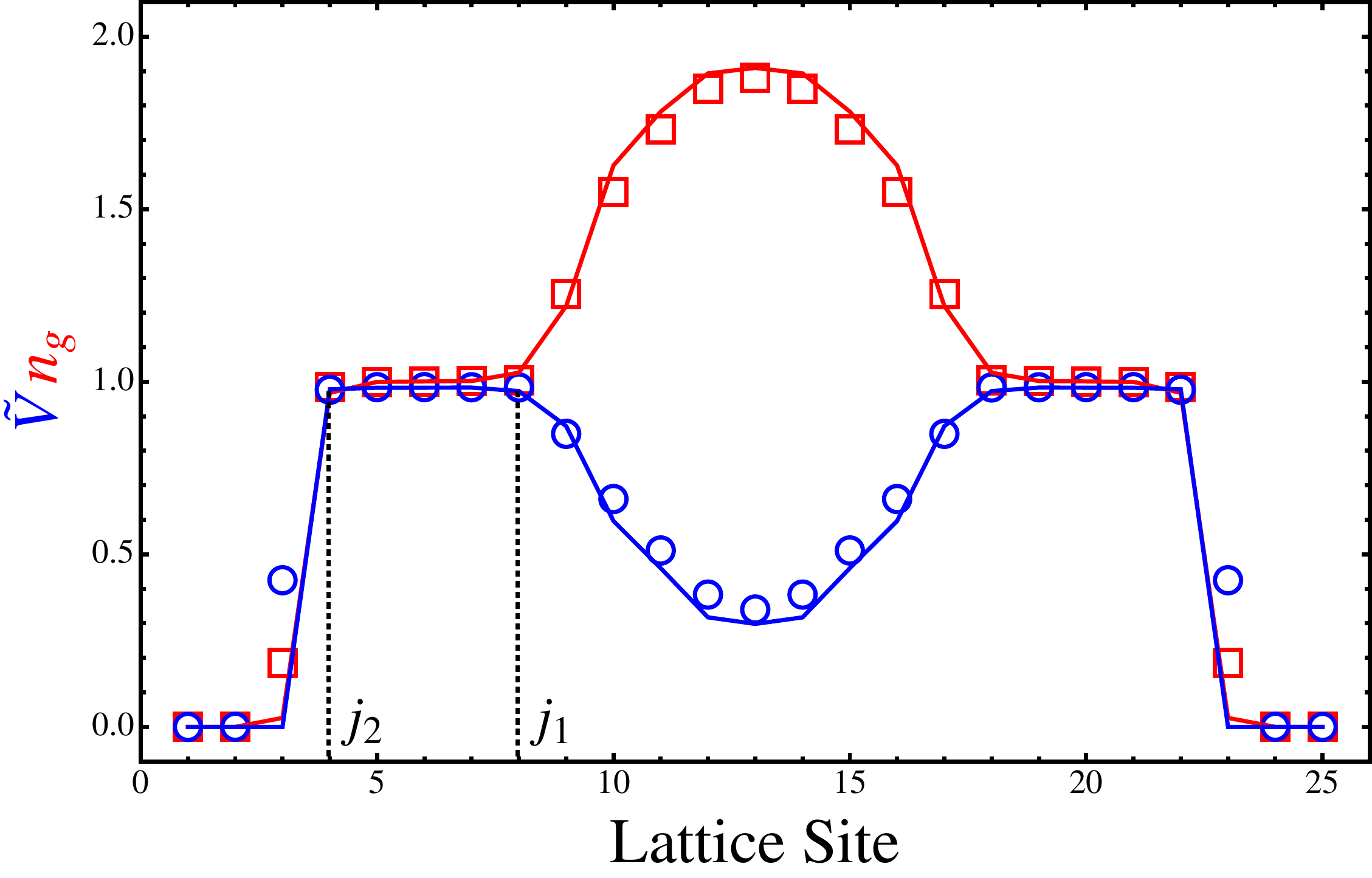}
}
\subfigure[][]{
\label{mf}
\includegraphics[height=3.5 cm]{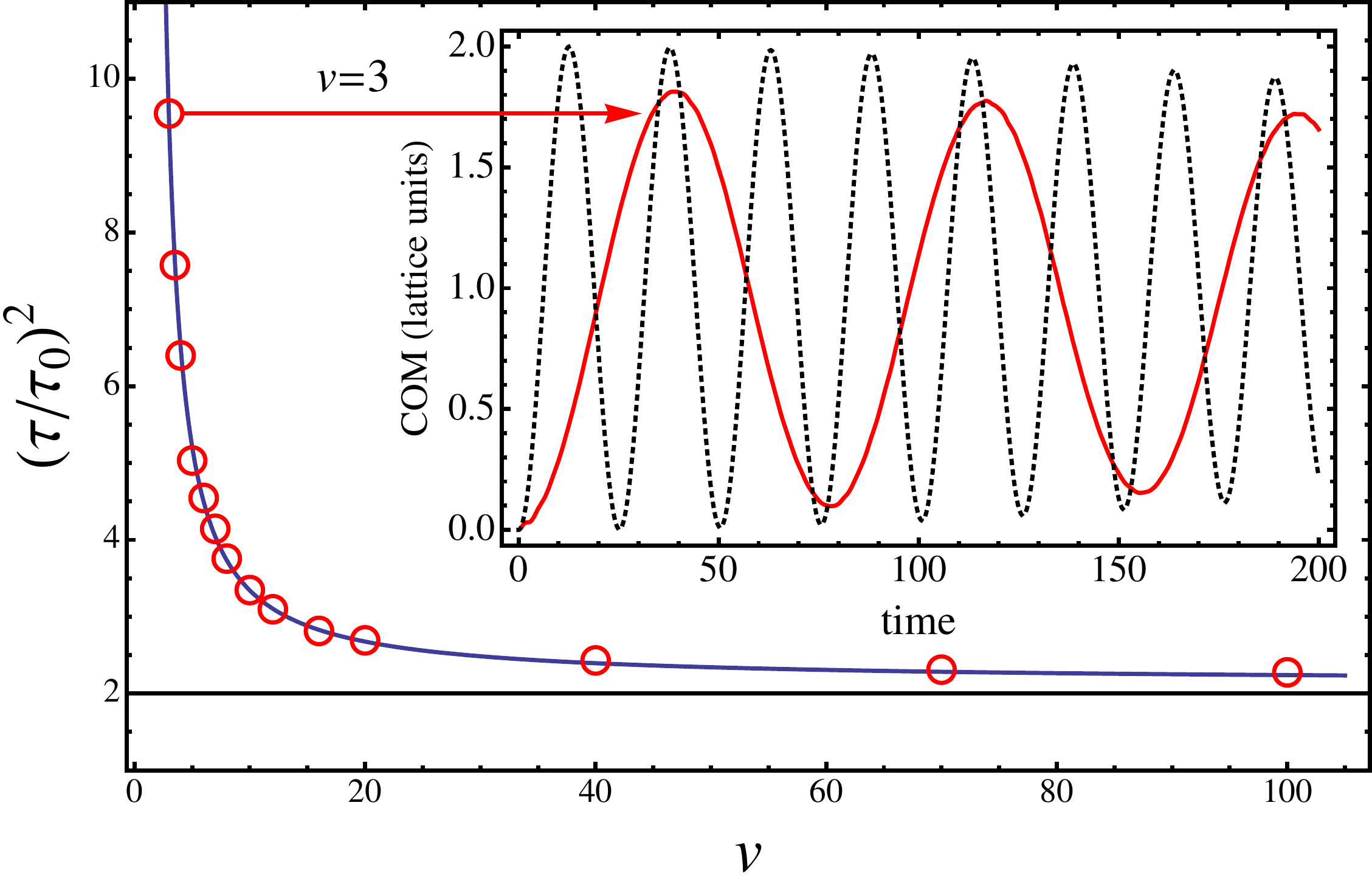}
}
\caption{(Color Online) \subref{phyb} In the translationally invariant MFT ($\Omega=0$), hybridization causes the crossing between the conduction and
  localized bands (black dotted lines) to be avoided, and opens a gap $\Delta_H$.  Fillings of $n_g<1$ correspond to
  filling only the lower band, and mass enhancement is due to placement of the Fermi Surface in the flattened band edges.  \subref{hybplateau} ($N_g=25,q=40,v=8$) The hybridization gap $\Delta_H$ induces a density plateau in the trap.  Lines are from self-consistent MFT and open shapes from LDA.  \subref{mf} ($N_g=5,q\approx 235$) Enhancement of the $g$ atom COM oscillation period ($\tau/\tau_0$) reveals the mass enhancement ($m/m_0\sim \tau^2/\tau_0^2$, with $m_0$ and $\tau_0$ the values at $v=0$).  Circles are from MFT dynamics, curve is a guide to the eye.  Inset: COM motion for $v=3$ (red line) and $v=0$ (black dotted line).}
\end{figure*}

If the $g$ atoms have negligible interactions with each other (true to very good approximation in $^{171}\textrm{Yb}$), and the strong repulsion between $e$ atoms is taken into account by a unit filling constraint, the low energy Hamiltonian contains only two interaction parameters $U_{eg}^{\pm}\propto a_{eg}^{\pm}\int d^3rw_{e}^2\left(r\right)w_{g}^2\left(r\right)$ ($w_{\alpha}$ is the lowest Wannier orbital for the lattice containing the $\alpha$ atoms).  Defining $V_{ex}=\left(U^+_{eg}-U^-_{eg}\right)/2$, dropping constant terms, and including a harmonic trap of curvature $\Omega$, the Hamiltonian reduces to \cite{Gorshkov:2009p4747}:
\begin{eqnarray}
\label{Hamiltonian}
\mathcal{H}_K&=&-J_g\sum_{\left<i,j\right>,\sigma}c^{\dagger}_{ig\sigma}c^{}_{jg\sigma}+V_{ex}\sum_{i\sigma\sigma'}c^{\dagger}_{ig\sigma}c^{\dagger}_{ie\sigma'}c^{}_{ig\sigma'}c^{}_{ie\sigma}\nonumber\\
&+&\Omega\sum_{i}i^2n_{ig}.
\end{eqnarray}
In the above $c^{\dagger}_{i\alpha\sigma}$ creates an
atom at site $i$ in electronic state $\alpha\in\left\{e,g\right\}$ and
(nuclear) spin state $\sigma$, $n_{i\alpha}\equiv\sum_{\sigma}c^{\dagger}_{i\alpha\sigma}
c_{i\alpha\sigma}$, and $J_g$ is the tunneling energy for the $g$ atoms.  The dimensionless ratios $v=-2V_{ex}/J_g$ and
$q=4J_g/\Omega$, together with the number of $g$ atoms $N_g$, characterize the different parameter regimes of the model.  It is important to note that both the sign and magnitude of $V_{ex}$ will depend on the atomic isotope used, and can be adjusted by offsetting one lattice from the other (to decrease the overlap between wanier orbitals) \cite{Gorshkov:2009p4747}.  Therefore, in principle both AF and FM exchange are relevant.  Nevertheless, in this paper we exclusively
consider the AF case ($v>0$).

In this regime the phase diagram of the translationally
invariant KLM in 1-D [Fig. \ref{PD1D}] has been relatively
well established, and can be drawn consistently from a variety of
numerical studies and several exact results \cite{Tsunetsugu:1997p597}. At strong coupling ferromagnetism prevails, but the weak coupling limit is paramagnetic (PM).  The boundary $n_g=1$ is insulating, having spin and charge gaps for arbitrarily small non-zero coupling.  To our knowledge the 1-D model is \emph{not} realized in condensed matter systems, but could be explored with AEAs in a 3-D optical lattice if both the $e$ and $g$ lattices were made deep in two of the dimensions (an array of 1-D tubes).

%
We begin our analysis in the PM phase, which is closely related to \emph{heavy fermion} behavior in
higher dimensions \cite{Tsunetsugu:1997p597}.  The mass enhancement can be
understood qualitatively through a hybridization mean field decoupling \cite{Lacroix:1979p981} in which the $g$ atoms near the Fermi surface gain large weight in the localized band [Fig. \ref{phyb}].  While the mean field theory (MFT) does not capture the Luttinger liquid nature of the PM phase at low energies, we believe it nonetheless provides a reasonable guide to the phenomena discussed here, and effects beyond MFT are left for future study.  Moreover, in work to be presented elsewhere, the calculations to follow have been extended to a 2D geometry where MFT is more reliable, with no qualitative change to the results.

The MFT can be obtained by a (non unique) decoupling of the interaction term in $\mathcal{H}_K$, leading to:
\begin{eqnarray}
\label{HMFT}
\mathcal{H}_{MFT}&=&-J_g\sum_{\left<i,j\right>\sigma}c^{\dagger}_{ig\sigma}c^{}_{jg\sigma}+\sum_{i\sigma}\left(\Omega i^2n_{gi}+\mu_i\left[n_{ei}-1\right]\right) \nonumber \\
&+&V_{ex}\sum_{i\sigma}\tilde{V}_i\left(c^{\dagger}_{ig\sigma}c_{ie\sigma}+h.c.\right)-V_{ex}\sum_{i}\tilde{V}_i^2
\end{eqnarray}
In Eq. (\ref{HMFT}) we have defined $\tilde{V}_i=\frac{1}{2}\sum_{\sigma}\left<c^{\dagger}_{ie\sigma}c_{ig\sigma}+h.c.\right>$, where the expectation value is taken in the slater determinant of the $(N_e+N_g)/2$ lowest energy single particle states (the $1/2$ accounts for spin degeneracy).  We have also introduced chemical potentials $\mu_i$ to enforce the local constraints $\langle n_{ie} \rangle=1$.  This decoupling is paramagnetic, and therefore cannot capture any magnetism, but it does describe the tendencies towards singlet formation at strong coupling.  In addition, it turns out to be the exact $N\rightarrow\infty$ solution of the $SU(N)$ generalization of the KLM (which can be implemented with AEAs having nuclear spin $I=\left(N-1\right)/2$) \cite{Coleman:2009p2107,Read:1984p2396}.
Beause $\mathcal{H}_{MFT}$ is quadratic it can be diagonalized, but it is necessary to choose the $\tilde{V}_i$ self consistently.

In the translationally invariant problem it is customary to assume $\tilde{V}_i = \tilde{V}$ and $\mu_i = \mu$, in which case analytic progress is possible.
With the trap we retain the site dependent $\tilde{V}_i$ and $\mu_i$, and self
consistent solutions must be obtained numerically.  The
procedure involves an  initial guess for the $\tilde{V}_i$ based
on Local Density Approximation (LDA): we treat the trap as a site dependent chemical potential, and infer the energy on each site from a translationally invariant problem.  LDA results are obtained by minimization of the energy thus obtained, while obeying a constraint on the total particle number.  We then solve for $\mu_i$ that
satisfy the local constraints \cite{Hermele}, diagonalize $\mathcal{H}_{MFT}$, and calculate the $\tilde{V}_i$.  Iterating this procedure we arrive at a self consistent solution.

From the MFT ground states we can easily compute the $\left<n_{ig}\right>$, which give us density profiles in the trap.  For $N_g$ or $\Omega$ sufficiently large these show plateaus [Fig \ref{hybplateau}] similar to what is observed for the repulsive Hubbard Model, although here they reflect the gap of a Kondo insulator, not a Mott insulator.  The Kondo insulator can be understood  from the hybridized band picture [Fig \ref{phyb}].  Unit filling of $g$ atoms correspongs to completely filling the lower band, and there is a charge gap of $\Delta_H$.  LDA considerations then imply that $\Omega\left(j_2^2-j_1^2\right)=\Delta_H$  [Fig. \ref{hybplateau}].  Exact results for the $v=\infty$ KLM give $\Omega\left(j_2^2-j_1^2\right)=3\left|V_{ex}\right|$; in this limit $\Delta_H$ tends to $2\left|V_{ex}\right|$, so the MFT underestimates the plateau size.  For the bosonic Hubbard model, where the relevant gap is the onsite interaction $U$, such plateau structures have already been imaged via microwave spectroscopy \cite{Campbell:2006p5171,Folling:2006p5135}.  We therefore expect that for large $v$ the plateau can be observable experimentally.

At lower fillings, such that the plateau does not form, we are everywhere in the heavy fermion metallic state.  Under these conditions we consider an experiment where the trap center is suddenly displaced, causing dipole oscillations of the $g$ atom center of mass (COM).  These type of experiments have been implemented in alkali atoms to study 1-D and 3-D transport of interacting bosons and fermions \cite{Fertig:2005p4167,Strohmaier2007,McKay2008} and used to probe different quantum many-body regimes in these systems.  We calculate these dynamics self consistently, starting with the MFT ground states and shifting the Hamiltonian.  If $\alpha^{\dagger}_{q\sigma}=\sum_{i}\left(u_{q}^{i}c^{\dagger}_{ig\sigma}+v_q^ic^{\dagger}_{ie\sigma}\right)$ create
the eigenstates of $\mathcal{H}_{MFT}$ then the following set of discrete Schr\"{o}dinger equations governs the evolution of the $u^i_q\left(t\right)$ and $v^i_q\left(t\right)$ after displacing the trap by $\delta$ lattice sites:
\begin{eqnarray}
-\imath\dot{u}_i^q&=&t\left(u_{i-1}^q+u_{i+1}^q\right)-\Omega \left(i-\delta\right)^2u_i^q-V_{ex}\tilde{V}_iv_i^q \nonumber \\
-\imath\dot{v}_i^q&=&-\Omega\left(i-\delta\right)^2v_i^q-\mu_iv_i^q-V_{ex}\tilde{V}_iu_i^q.
\end{eqnarray}
When integrating these equations the $\tilde{V}_i$ are updated self consistently, and the $\mu_i$ are evolved in time by ensuring that $\ddot{n}_{ei}=0$ (the first time derivative has no dependence on the $\mu_i$).  Such time dependence of the $\mu_i$ preserves the one  $e$ atom per site constraint, and its necessity has a simple origin: $\mathcal{H}_{MFT}$ breaks the local $U(1)$ symmetry of $H_K$ associated with conservation of the $e$ atom density.

The $g$ atom COM oscillations ensuing from displacement of the trap by one lattice
site have been obtained at several different values of $v$
for fixed $q$. We find a strong enhancement of the oscillation period $\tau$ (and hence of the quasiparticle mass $m\sim\tau^2$) for decreasing $\left|V_{ex}\right|$ [Fig. \ref{mf}].  Once $v\sim 1$, the comparably fast non-interacting oscillations emerge on top of the slow oscillations of the heavy quasiparticles, converging to the noninteracting result as $v \to 0$ -- this cuts off the apparent trend toward diverging $\tau$ in Fig.~\ref{mf}.



\begin{figure}[h]
\centering
\subfiguretopcaptrue
\subfigure[][]{
\hspace{-.1in}
\label{free}
\includegraphics[height=2.5 cm]{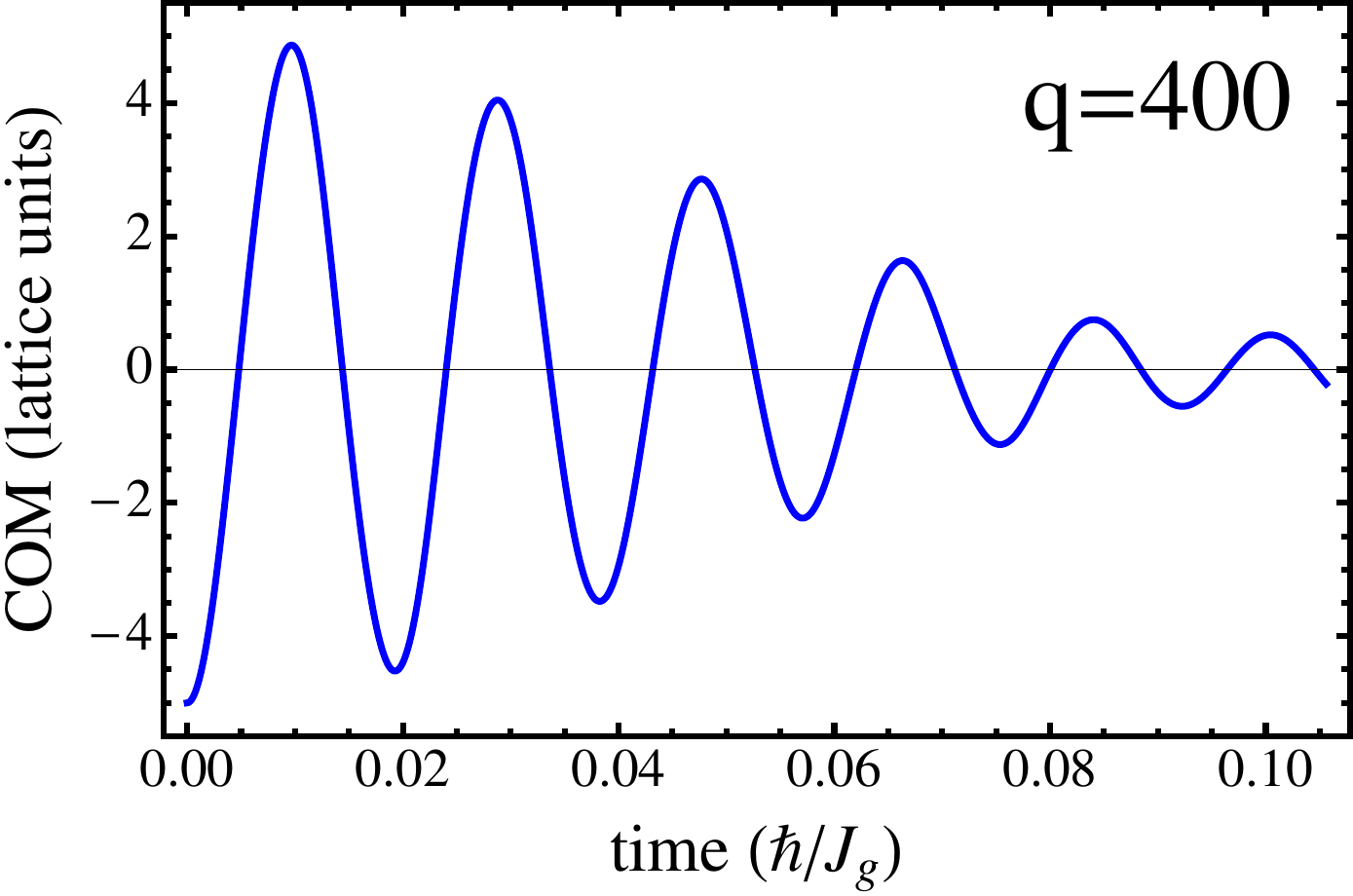}
}
\hspace{-.05in}
\subfigure[][]{
\label{fermionized}
\includegraphics[height=2.5 cm]{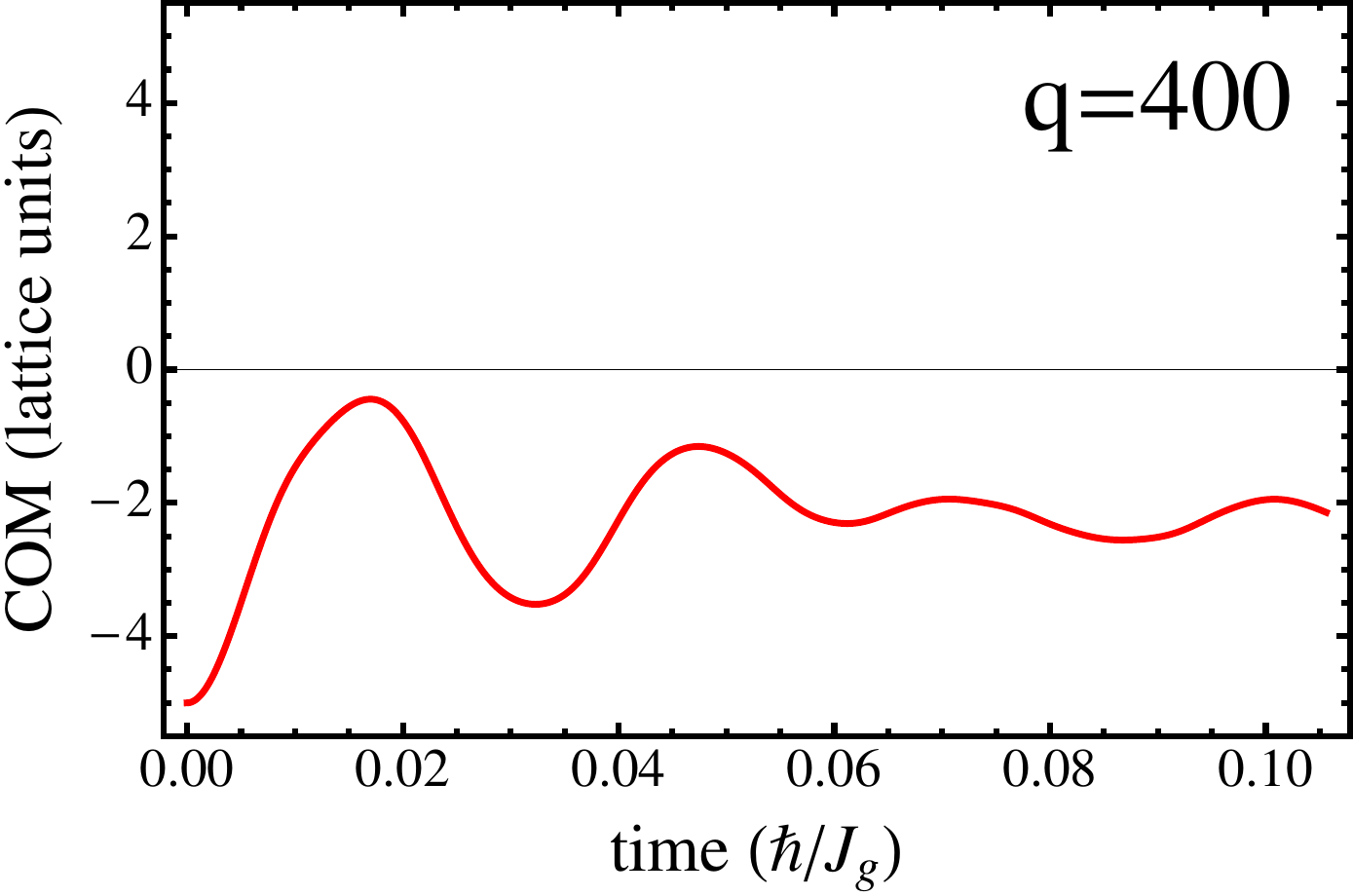}
}
\caption{(Color Online) \subref{free} Non interacting dynamics of $16$ $g$ atoms after displacement of the trap by $5$ lattice sites.  \subref{fermionized} Same as \subref{free}, but now with $v=\infty$ (calculated using $\mathcal{H}_{\infty}$).}
\label{StrongCouplingFig}
\end{figure}


We now turn to the FM part of the phase diagram, which exists for all fillings at sufficiently strong coupling.  When $v=\infty$ the ground state is formed by pairing each $g$ atom into a spin singlet with one $e$ atom [Fig. \ref{phasediagrams}(b)].  To first order in $J_g$ the singlets become mobile, and there is an exact mapping of the unpaired $e$ atoms to the fermions of a $U=\infty$ Hubbard model (the singlets are the holes) \cite{Lacroix:1985p1539}. Because nearest neighbor hopping cannot exchange up and down spins, we can think of the fermions as spinless \cite{Ogata:1990p1596}, perform a particle hole transformation (now the singlets are the spinless fermions), and thereby arrive at a simple Hamiltonian describing the $g$ atoms:
\begin{equation}
\label{strongcoupling}
\mathcal{H}_{\infty}=-\frac{J_g}{2}\sum_{\left<i,j\right>}c^{\dagger}_{ig}c^{}_{jg}+\Omega\sum_{i}i^2n_{ig}.
\end{equation}
The reduction of the hopping energy is the result of projecting out the high energy triplet states.  To highlight this strong coupling behavior, we again consider dynamics ensuing from a sudden displacement of the trap center.  If the system is strongly interacting, Eq. (\ref{strongcoupling}) avails a simple treatment of these dynamics based on the non-interacting solutions in Ref. \cite{Rey:2005p141}.  There the authors found that for $q\gg1$ and $N_g\lesssim 4\sqrt{2J_g/\Omega}$ the dynamics involved \emph{delocalized}, free space harmonic oscillator like states with level spacing $\omega^*=\Omega\sqrt{q}$.  At $v=\infty$ we effectively have $N_g\rightarrow2N_g$ (because the fermions become spinless) and $J_g\rightarrow J_g/2$, therefore the inequality can be violated at strong coupling even when satisfied for the non-interacting system; \emph{localized} states become populated, and transport is strongly inhibited (see Fig. \ref{StrongCouplingFig}).

\begin{figure}[h]
\centering
\subfiguretopcaptrue
\subfigure[][]{
\hspace{-.1in}
\label{spectrum}
\includegraphics[height=2.5 cm]{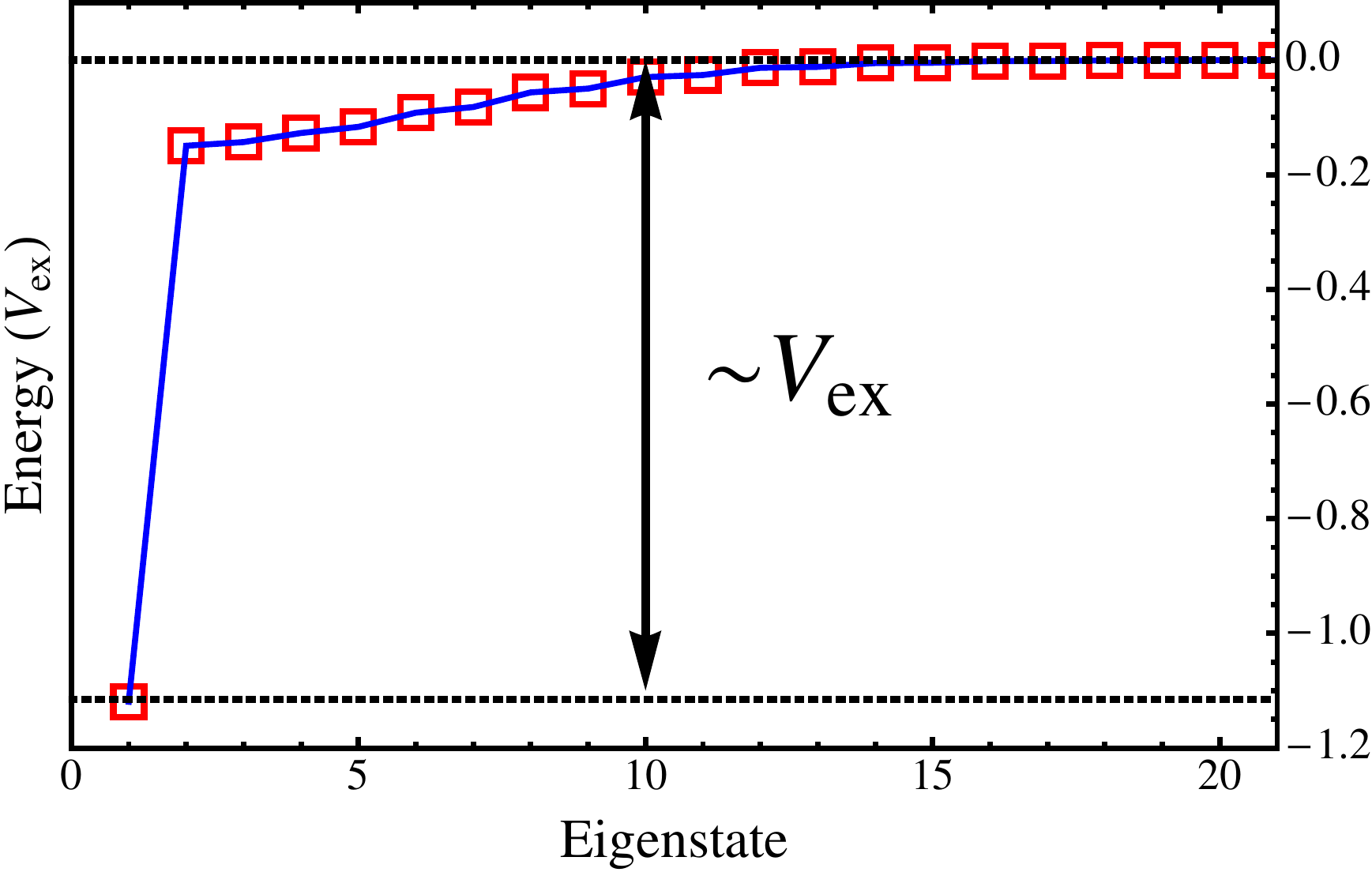}
}
\hspace{-.05in}
\subfigure[][]{
\label{poscill}
\includegraphics[height=2.5 cm]{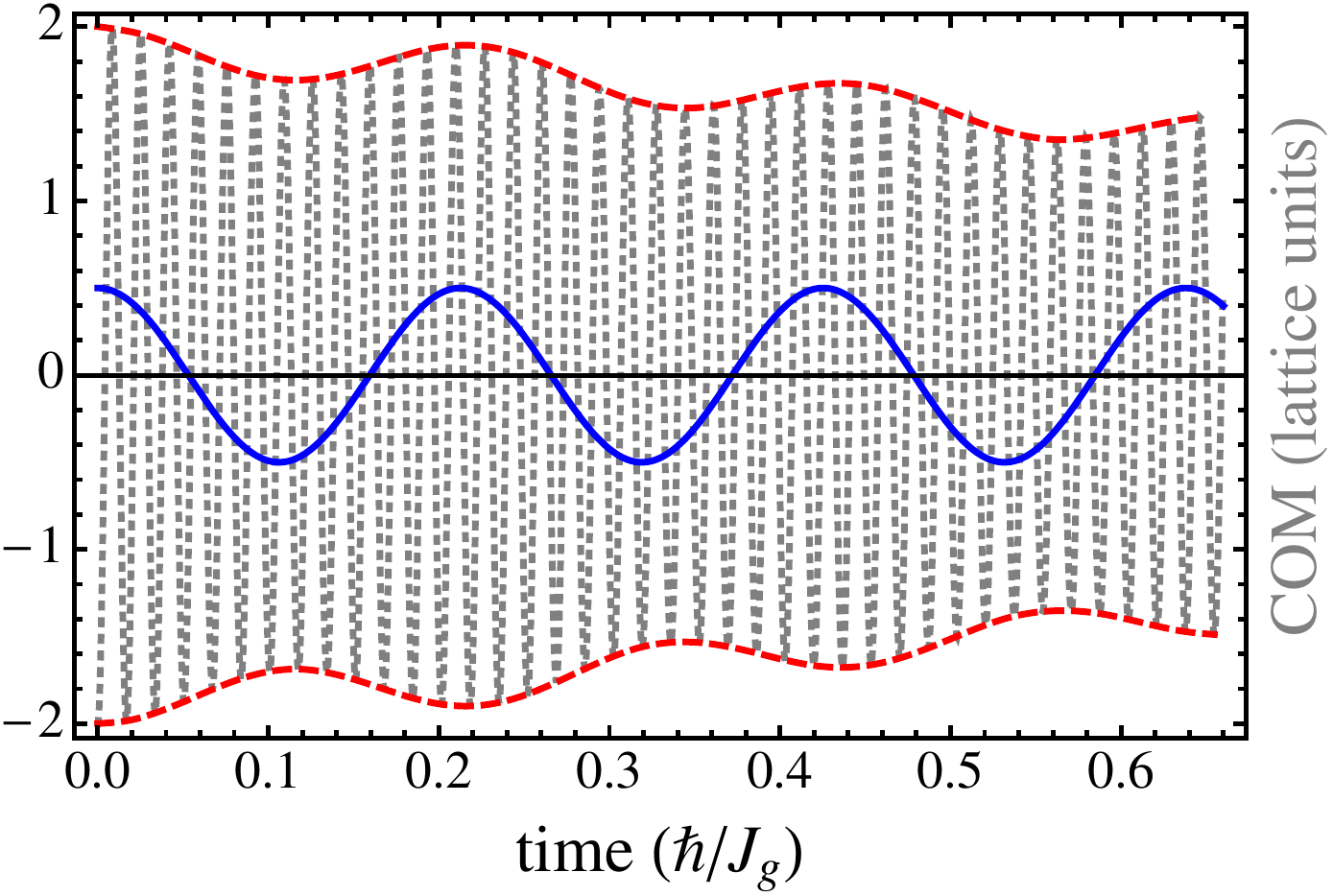}
}
\caption{(Color Online) \subref{spectrum} Low energy polaron spectrum for $q=800$, showing a gap of size $\sim V_{ex}$.  \subref{poscill} ($q=800, v=0.02$) Polaron COM (black dotted) oscillating after trap displacement by $2$ lattice sites.  For comparison we plot $\cos(V_{ex}t)$ (blue solid line), showing that the energy gap determines the time scale of amplitude modulation.  The overall decay is a finite size effect, and eventually revives.}
\label{pics}
\end{figure}

Another limit which is well understood in the FM phase of the translationally invariant model is $N_g=1$.  Sigrist et al. \cite{Sigrist:1991p5100} proved that
the ground state of the KLM with $L$ sites and one conduction electron has
total spin $S=\frac{1}{2}\left(L-1\right)$ even in the absence of
translational symmetry, and they described the excitations for the homogenous case as bound states between the $g$ atom and a flipped spin in the deep lattice (spin polarons, see Fig. \ref{phasediagrams}(b)).  For a weak trap ($q\gg 1$) and at sufficiently small coupling ($V_{ex} \ll\omega^*$), we characterize the polaron spectrum to lowest order in degenerate perturbation theory and find that
one eigenvalue separates from the rest by a gap of approximately
$\left|V_{ex}\right|$ [Fig. \ref{spectrum}].  As seen in Fig. \ref{poscill}, this energy scale is manifest in the COM oscillations of a single $g$ atom as strong modulation of the
oscillation amplitude
with periodicity $\tau\approx\frac{2\pi}{\left|V_{ex}\right|}$, as verified by dynamics calculated from the exact eigenstates.  

For the finite system under consideration, we expect this behavior will persist for $N_g>1$ whenever $N_g$ is odd.  This can be seen by noting that a single unpaired $g$ atom at the Fermi level gains energy at first order in perturbation theory when coupling to the $e$ atoms is turned on, whereas the doubly occupied levels below it do so only at second order.  It is worth noting that the condition $V_{ex}\ll\omega^*$ is equivalent to demanding the perturbation stay smaller than the finite size gap.  For fixed lattice depth and peak $g$ atom density, the gap scaling is $\omega^*\sim1/R$, with $R$ the Thomas Fermi radius of the $g$ atom cloud ($R\propto\sqrt{N_gJ_g/\omega^*}$).  This means that the demonstrated modulations will be washed out with increasing $\left|V_{ex}\right|$ or with increasing $g$ atom number, and are manifestly a finite size effect.




We now consider the feasibility of generating and observing these dynamics in an experiment.  Throughout the paper we assume a unit filled Mott insultor of $e$ atoms coexisting with various fillings of $g$ atoms at the center of a trap.  To realize this situation in experiment, one could  first ramp up a deep optical lattice for the $g$ atoms such that they exhibit a Mott insulator shell structure. By taking advantage of the energy shift between double and single occupied sites,  it is then possible  to selectively transfer atoms into the $e$ state, in such a way that sites with two $gg$ atoms become $eg$ sites and single occupied $g$ sites become single occupied $e$ sites.  A subsequent  adiabatic reduction of the lattice depth for the $g$ atoms achieves the desired configuration.  It may also be helpful to confine the $g$ atoms more tightly than the $e$ atoms (to ensure that they do not sample the wings of the $e$ atom Mott insulator), which can be achieved by blue detuning the deep lattice.  In most of our calculations we have used small trap displacements to simplify the numerics, and in a real experiment they will inevitably be larger.  Observation of the dipole oscillations in 1-D with amplitude of $\lesssim 8$ lattice sites has precedent, and was carried out by mapping the center of mass position of the atomic cloud to momentum space\cite{Fertig:2005p4167,Strohmaier2007}.  Moreover, a recent proposal \cite{Peden:2009p5704} suggests that dynamics could be characterized from a single non destructive measurement if the atoms are coupled to an unpumped cavity field.  The small atom numbers necessary for observation of the modulations at small $\left|V_{ex}\right|$ also has precedent, with $\sim18$ atoms per tube having been realized in an array of 1D lattices \cite{bloch}.  We emphasize that all physics discussed in the paper involves energy scales on the order of $V_{ex}$, which makes temperature demands less constraining than for proposals involving superexchange or RKKY like physics ($v\approx 1$ gives a Kondo temperature on the order of $\left| V_{ex}\right| / k_B$, so this statement applies even to the heavy fermion behavior).

The results presented demonstrate that dipole oscillations of the $g$ atom COM effectively probe a variety of KLM phenomena.  The emphasis has been on a 1D system, which is of relevance to cold atom experiments, primarily because current theoretical understanding of the phase diagram is strongest here.  However, the heavy fermion behavior generalizes to experiments in 2 and 3 dimensions. Though it has not been discussed, we point out that an optical lattice experiment---especially in $D>1$---is a natural setting for probing the size of the Fermi surface in the heavy fermion state, and could corroborate evidence for a large Fermi surface observed in condensed matter experiments.

We thank Thomas Gasenzer, Matthias Kronenwett, Alexey Gorshkov, Maria Luisa Chiofalo, Brandon Peden, Victor Gurarie and Jun Ye for helpful discussions.  This work was supported by grants from the NSF and from DARPA (OLE Program).

\bibliography{SP1v8.bib}

\end{document}